\documentclass[superscriptaddress,showpacs,floatfix,preprintnumbers, nofootinbib,hyperref]{revtex4} 
\usepackage{graphicx}
\usepackage{rotating}
\usepackage{epsfig}
\usepackage[usenames,dvipsnames]{xcolor}

\usepackage{color}

\def\he4{$^4$He}
\def\h2{$^2$H}

\newcommand{\lesssim}{\,\rlap{\lower3.7pt\hbox{$\mathchar\sim$}}
\raise1pt\hbox{$<$}\,}

\begin{document}

\title{The strongest bounds on active-sterile neutrino mixing after Planck data\footnote{Based on observations obtained with Planck, an ESA science mission with instruments and contributions directly funded by ESA Member States, NASA, and Canada, (http://www.esa.int/Planck).}}

\author{Alessandro Mirizzi} 
\affiliation{II Institut f\"ur Theoretische Physik, Universit\"at Hamburg, Luruper Chaussee 149, 22761 Hamburg, Germany}
\author{Gianpiero Mangano}
\affiliation{Istituto Nazionale di Fisica Nucleare - Sezione di Napoli, Complesso Universitario di Monte S. Angelo, I-80126 Napoli, Italy}
\author{Ninetta Saviano} 
\affiliation{II Institut f\"ur Theoretische Physik, Universit\"at Hamburg, Luruper Chaussee 149, 22761 Hamburg, Germany} 
\affiliation{Istituto Nazionale di Fisica Nucleare - Sezione di Napoli, Complesso Universitario di Monte S. Angelo, I-80126 Napoli, Italy}
 \affiliation{Dipartimento di Fisica, Universit{\`a} di Napoli Federico II, Complesso Universitario di Monte S. Angelo, I-80126 Napoli, Italy}
\author{Enrico~Borriello} 
\affiliation{II Institut f\"ur Theoretische Physik, Universit\"at Hamburg, Luruper Chaussee 149, 22761 Hamburg, Germany}
\author{Carlo Giunti}
\affiliation{Department of Physics, University of Torino and INFN, Via P. Giuria 1, I-10125 Torino, Italy}
\author{Gennaro Miele}
\affiliation{Istituto Nazionale di Fisica Nucleare - Sezione di Napoli, Complesso Universitario di Monte S. Angelo, I-80126 Napoli, Italy}
\affiliation{Dipartimento di Fisica, Universit{\`a} di Napoli Federico II, Complesso Universitario di Monte S. Angelo, I-80126 Napoli, Italy}
 \author{Ofelia Pisanti}
 \affiliation{Istituto Nazionale di Fisica Nucleare - Sezione di Napoli, Complesso Universitario di Monte S. Angelo, I-80126 Napoli, Italy}
 \affiliation{Dipartimento di Fisica, Universit{\`a} di Napoli Federico II, Complesso Universitario di Monte S. Angelo, I-80126 Napoli, Italy}


\begin{abstract}
Light sterile neutrinos can  be excited by oscillations with active neutrinos in the early universe.
Their properties can be constrained by their contribution as  extra-radiation, parameterized in terms of the effective number of neutrino species $N_{\rm eff}$, 
and to the universe energy density today $\Omega_\nu h^2$. Both these parameters have been measured to quite a good precision by the Planck satellite experiment. 
 We use this  result to update the bounds on the parameter space of  (3+1) sterile neutrino scenarios, with an active-sterile neutrino mass squared splitting in the range $(10^{-5} - 10^2 )\,\textrm{eV}^2$.
We consider  both normal and inverted mass orderings for the active and sterile states. For the first time we take into account the possibility of two non-vanishing active-sterile mixing
angles.
We find that the bounds  are more stringent than those obtained in laboratory experiments. 
This  leads to  a strong tension with the short-baseline hints of light sterile neutrinos.
In order to relieve this disagreement, modifications of the standard cosmological scenario, e.g. large primordial
neutrino asymmetries, are required.
\end{abstract}

\pacs{14.60.St, 
	   14.60.Pq, 
		98.80.-k 
		26.35.+c 
}  

\maketitle

\section{Introduction}  
    Light sterile neutrinos have been advocated as a possible solution to some puzzling results found  in neutrino oscillations, see~\cite{Abazajian:2012ys} for a recent review.
In particular,  $m\sim\mathcal O(1)\,$eV sterile neutrinos mixing with the active states have been proposed to solve different
anomalies observed in short-baseline neutrino experiments,
in the ${\overline \nu}_\mu \to {\overline\nu}_e$ oscillations in
LSND~\cite{Aguilar:2001ty} and MiniBoone~\cite{AguilarArevalo:2010wv,Aguilar-Arevalo:2013pmq} experiments, and in ${\overline\nu}_e$ and $\nu_e$ disappearance
revealed by the Reactor Anomaly~\cite{Mention:2011rk} and the Gallium Anomaly~\cite{Acero:2007su}, respectively. 
Scenarios with one (dubbed ``3+1'') or two (``3+2'') sub-eV sterile neutrinos~\cite{Giunti:2012tn,Kopp:2013vaa,Conrad:2012qt} have been proposed
to fit the different data. 
On the other hand, lighter sterile neutrinos with $\Delta m^2 \sim 10^{-5}\,$eV$^{2}$ can explain the absence of the upturn 
in the solar neutrino energy spectrum~\cite{deHolanda:2010am}.

The main theoretical motivations for these states are perhaps not so strong, though 
 light sterile neutrinos with a sizable mixing 
emerge in several models (see, e.g.,~\cite{Merle:2013gea,Duerr:2013opa} and references therein). In any case, 
since their discovery would point towards new physics, their role is relevant enough to justify an open mind attitude and a close looking for any, yet tiny, evidence for new effects beyond the {\it too much} successful Standard Model.
 
 The hunt for sterile neutrinos in laboratory experiments is currently open. Different  techniques
have been proposed to search for these particles
 (see, e.g.,~\cite{Abazajian:2012ys,Kose:2013zsa,Rubbia:2013ywa}). Indeed,  
 since every experimental measurement  has its own systematic uncertainties and its own recognized or un-recognized
loop holes, 
in order to corner the  sterile neutrino parameter space  it is  worth using as many  
handles as possible (see, e.g.,~\cite{Palazzo:2011rj,Wu:2013gxa}).  In this respect, cosmology is one of the most powerful
tools (see,
 e.g.,~\cite{Hannestad:2003ye,Cirelli:2004cz}). Adding exotic contribution to the radiation content in the universe, usually expressed in terms of the effective number of excited neutrino species, $N_{\rm eff}$, has in fact, a big impact on both the Cosmic Microwave Background (CMB) anisotropy map \cite{Komatsu:2010fb,Hinshaw:2012fq,Hou:2012xq,Sievers:2013wk}, and the Big Bang Nucleosynthesis (BBN) nuclear species yields \cite{Mangano:2011ar,Hamann:2011ge}.
The standard expectation for this parameter is $N_{\rm eff}=3.046$~\cite{Mangano:2005cc}.
If low-mass sterile neutrinos exist and mix with active flavors, they can be thermally excited by the interplay of
oscillations and collisions, producing a larger value of $N_{\rm eff}$. Cosmological constraints on sterile neutrinos based on their contribution to the extra-radiation have been presented in several papers, see e.g.~\cite{Dolgov:2003sg,Cirelli:2004cz,Chu:2006ua,Hannestad:2012ky}.

In the last few years, a possible cosmological hint in favor of light sterile neutrinos (see e.g.~\cite{Hamann:2010bk,GonzalezGarcia:2010un}) was found by combining the result in the  best fit of WMAP, SDSS II-Baryon Acoustic Oscillations and Hubble Space Telescope (HST) data, yielding a $68~\%$ C.L. range on $N_{\rm eff}=4.34^{+0.86}_{-0.88}$~\cite{Komatsu:2010fb} for a $\Lambda$CDM universe.
The recent results of WMAP-9~\cite{Hinshaw:2012fq}, SPT \cite{Hou:2012xq} and ACT \cite{Sievers:2013wk}, exploiting the damping tail features at high multipoles, have weakened this evidence to less than 2-$\sigma$.

A recent breakthrough in constraining the {\it dark} radiation content in the early universe is represented by the first data release of the Planck experiment \cite{Planck2013}, a satellite  with unprecedented sensitivity in the high multipole range.
Indeed, one of the main result of this new CMB anisotropy map is the quite accurate estimate of the relativistic degrees of freedom at recombination epoch, $N_{\rm eff} = 3.30 \pm 0.27$ at 68~\% C.L., a result obtained combining Planck, WMAP, Baryon Acoustic Oscillation and high multipole CMB data \cite{Planck2013}. Within 1-$\sigma$ this is compatible with the standard expectation, but still leaves room for almost an extra neutrino species at 95~\% C.L. 
Moreover, combining the Planck data with the Hubble constant $H_0$ measurement from HST, the best-fit increases to $N_{\rm eff} = 3.62 \pm 0.25$ at 68~\% C.L. This would amount to a $2.3$-$\sigma$ signal for extra-radiation, and different models producing a moderate amount of extra-radiation have been proposed 
(see, e.g.,~\cite{Cicoli:2012aq,DiBari:2013dna,DiValentino:2013qma,Said:2013hta,Franca:2013zxa,Conlon:2013isa}). 
However,  there is clearly a tension 
between the   Planck and HST determination of $H_0$ in the $\Lambda$CDM model, and at the moment  possible systematic effects in the astrophysical determination of $H_0$ cannot be excluded. Therefore, in the following
we will present our bounds on sterile neutrinos, using the determination of $N_{\rm eff}$ without the inclusion of HST data.

The measured value of $N_{\rm eff}$ is not the only cosmological parameter  that can be used to constrain 
massive neutrinos. Indeed, 
above $m \sim {\mathcal O} (1)$~eV, sterile neutrinos become non-relativistic at the CMB epoch. Therefore, their contribution
to the standard radiation becomes sub-dominant for larger masses~\cite{Jacques:2013xr}. However,  
  they contribute to the energy density  in the Universe today,
$\Omega_\nu h^2$, which for fully thermalized non-relativistic neutrinos, is directly proportional to their number density, 
i.e.
\begin{equation}
\Omega_\nu h^2 = \frac{\sum m_\nu}{94.1 \,\ \textrm{eV}} \,\ ,
\end{equation}
where $h$ is the Hubble constant in units of 100 km s$^{-1}$ Mpc$^{-1}$. 
Planck analysis provides also joint constraints on $N_{\rm eff}$ and ${\sum m_\nu}$
in different models for active and sterile neutrinos.
In particular, assuming the existence of a thermalized massive sterile neutrino together  with two massless
active neutrinos and a massive one with mass fixed by the atmospheric mass
splitting (i.e. $m \sim 0.06$~eV), one would get as  bound combining Planck, WMAP, Baryon Acoustic Oscillation and high multipole CMB data,
\begin{eqnarray}
N_{\rm eff} &<& 3.80 \,\ , \nonumber\\
m^{\rm eff}_{s} &<& 0.42 \,\ \textrm{eV} \,\ ,
\label{planckneff}
\end{eqnarray}
at  95~\% C.L. (see Eq.~(83) of Ref.~\cite{Planck2013}), where the effective sterile neutrino mass is 
$m^{\rm eff}_{s} = 94.1 \times \Omega_\nu h^2$~eV. This mass bound strengthens the previous one obtained 
in~\cite{Hamann:2010bk} using  WMAP, SDSS II-Baryon Acoustic Oscillations and Hubble Space Telescope (HST) 
data.

These new cosmological data are the motivation of this paper, where we present an update of the cosmological bounds on light sterile neutrinos. We focus on (3+1) scenarios, considering a broad range for active-sterile neutrino mass splitting, which cover the regions where laboratory  hints
 emerge. We have chosen to consider a minimal scenario (one extra sterile state only), and not to cover the (3+2) case, where it seems harder to fit the neutrino mass bound from large scale structure~\cite{Hamann:2010bk,Melchiorri:2008gq,Archidiacono:2013xxa}.  This case also appears to be disfavored by Planck results unless one considers large neutrino asymmetries to suppress the sterile neutrino production, see e.g. \cite{Chu:2006ua,Hannestad:2012ky,Saviano:2013ktj}. 
 
 The outline of our work is as follows. In Sec.~II we present the setup of the flavor evolution in the (3+1) scenarios we are considering. In Sec.~III we  show the bound on the sterile neutrino parameter space coming from the extra-radiation
content $N_{\rm eff}$ and from the energy density $\Omega_\nu h^2$. Finally, in Sec.~IV we summarize our results and we conclude.

\begin{figure*}
\begin{center}
 \includegraphics[width=0.8\textwidth]{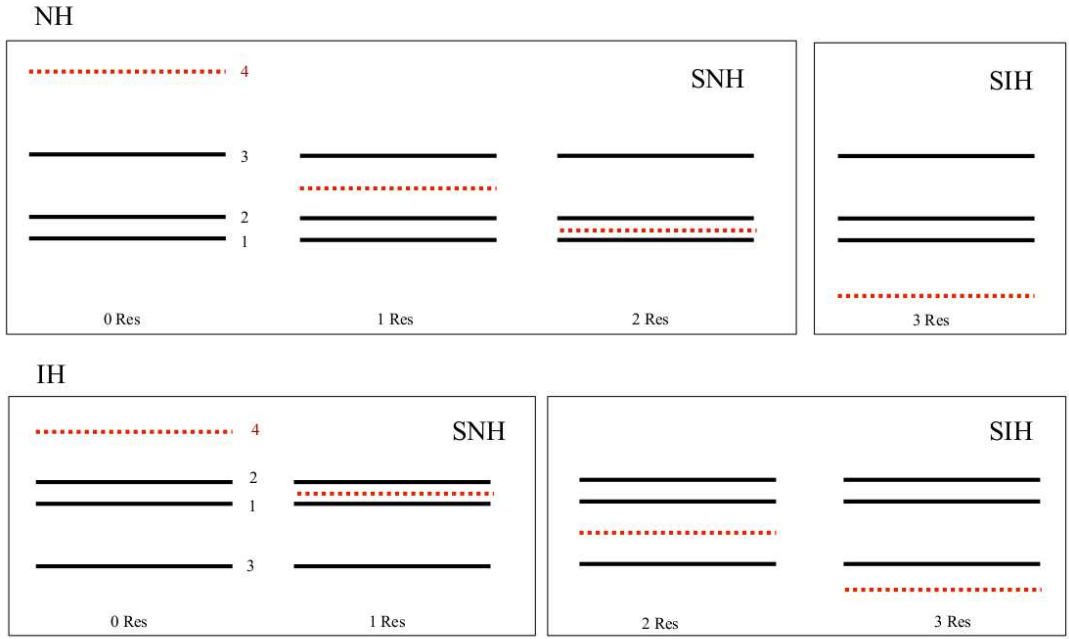} 
 \end{center}
\caption{Active and sterile neutrinos mass orderings with the  scheme of possible resonances.} \label{fighier}
\end{figure*}

 \section{Setup of the  flavor evolution}

\subsection{(3+1) mixing schemes}  
  
 We consider one single light sterile neutrino $\nu_s$, which 
mixes with the active neutrino states $\nu_e, \nu_{\mu}, \nu_{\tau}$.  The flavor eigenstates $\nu_\alpha$ are related to the mass eigenstates $\nu_i$  via a unitary matrix ${\mathcal U}=
{\mathcal U} (\theta_{12}, \theta_{13}, \theta_{23}, \theta_{14}, 
\theta_{24}, \theta_{34})$~\cite{Maltoni:2001bc,Abazajian:2012ys}
where we order the flavor eigenstates in such a way that if all mixing angles are vanishing we have $(\nu_e,\nu_\mu,\nu_\tau ,\nu_s)=(\nu_1,\nu_2,\nu_3, \nu_4)$. 
In the following we fix the values of the three active mixing angles to the current best-fit from global analysis of the different active neutrino oscillation data~\cite{Fogli:2012ua},
$\sin^2 \theta_{12} = 0.307$, $ \sin^2 \theta_{23} = 0.398$, and 
$\sin^2  \theta_{13} = 0.0245$. 
Concerning the mixing angles between active and sterile neutrinos we choose as representative range 
$10^{-5} \leq \sin^2 \theta_{i4} \leq 10^{-1}$  (i=1,2,3).

The 4$\nu$ mass spectrum is parameterized as
${\mathcal M}^2 =\textrm{diag}\left(m_1^2, m_1^2+
 \Delta m^2_{21} \, , 
   m_1^2+\Delta m^2_{31}  \, , m_1^2+\Delta m^2_{41}  \right)$.
   We  consider a hierarchical mass spectrum, obtained setting $m_1=0$. 
   This is consistent with the scenario considered by Planck, to obtain the constraint on the 
   sterile neutrino mass.
The solar and the atmospheric mass-square differences are given  by
$\Delta m^2_{21}= m_2^2-m_1^2= 7.54 \times 10^{-5}$~eV$^2$ and  
$|\Delta m^2_{31}| = |m_3^2-m_1^2|= 2.43 \times 10^{-3}$~eV$^2$, respectively~\cite{Fogli:2012ua}.
Depending on the sign of $\Delta m^2_{31}$ and $\Delta m^2_{41}$ we define an active normal mass hierarchy (NH, 
$\Delta m^2 _{31} >0$) or 
an active inverted mass hierarchy (IH, $\Delta m^2 _{31} <0$) and a sterile normal mass hierarchy
(SNH, $\Delta m^2 _{41} >0$) or a sterile inverted mass hierarchy (SIH, $\Delta m^2 _{41} <0$).
In our study we consider the following ranges  
$10^{-5} \leq \Delta m^2_{41}/\textrm{eV}^2 \leq 10^2$  in SNH and 
$10^{-5} \leq |\Delta m^2_{41}|/\textrm{eV}^2 \leq 10^{-2} $  in SIH.
Note that in SIH larger values of $|\Delta m^2_{41}|$ are disfavored  due to 
the cosmological bounds on the neutrino masses~\cite{Hinshaw:2012fq,Planck2013}.
The different combinations of the active and sterile mass orderings are shown in Fig.~\ref{fighier}. 

\subsection{Active-sterile neutrinos kinetic equations} 

The neutrino (antineutrino) ensemble in a medium, as in the early universe plasma, is described in terms of  a $4\times 4$ momentum-dependent density matrix $\varrho_{\bf p}$ ($\bar \varrho_{\bf p}$).
To solve the full  set of momentum dependent equations of motion~\cite{Sigl:1992fn} turns out to be a computationally demanding task (see, e.g.,~\cite{Hannestad:2012ky,Saviano:2013ktj} for  recent studies). 
Since our aim is to perform an extensive scan of the sterile neutrino parameter space,
 in order to carry out a more treatable numerical analysis,  we will consider the averaged-momentum approximation, based on the ansatz,
$ \varrho_{\bf p} (T) \to f_{FD}(p)\,\rho(T)\, $ (see~\cite{Mirizzi:2012we})
where $\rho(T)$ is  the density matrix for the mean thermal momentum $\langle p \rangle = 3.15~T$, and
 $f_{FD}(p)$ is the Fermi-Dirac neutrino equilibrium distribution, and similarly for antineutrinos. 

The evolution equation  for  the momentum-averaged density matrix $\rho$, describing the neutrino  system,  is the following~\cite{Sigl:1992fn,Dolgov:2002ab,Mirizzi:2012we}:
\begin{equation}
{\rm i}\,\frac{d\rho}{dt} =[{\sf\Omega},\rho]+ C[\rho]\,,
\label{drhodt}
\end{equation}
and a similar expression holds for the antineutrino matrix $\bar\rho$.
The evolution in terms of the comoving observer proper time $t$ can be easily recast in function of the temperature $T$ 
(see~\cite{Dolgov:2002ab} for a detailed treatment).
The first term on the right-hand side of Eq.\ (\ref{drhodt}) describes the flavor oscillations Hamiltonian,
\begin{equation}
{\sf\Omega}=\frac{{\sf M}^2}{2} \left\langle \frac{1}{p} \right\rangle +
\sqrt{2}\,G_{\rm F}\left[-\frac{8p}{3 }\, \bigg(\frac{{\sf E_\ell}}{m_{\rm W}^2} + \frac{{\sf E_\nu}}{m_{\rm Z}^2}\bigg)
+{\sf N_\nu}\right]\,,
\end{equation}
where ${\sf M}^2= {\mathcal U}^{\dagger} {\mathcal M}^2 {\mathcal U}$ is the neutrino mass matrix, while
the terms proportional to the Fermi constant $G_F$  encode the matter effects in the neutrino oscillations.
In particular, the term ${\sf E_\ell}$ is related to the energy density of $e^{\pm}$ pairs, 
${\sf E_\nu}$ to the energy density of $\nu$ and  $\bar\nu$,  and ${\sf N_\nu}$ 
is the $\nu-\nu$ interaction term  proportional to the neutrino asymmetry.
In the following, we will consider the most conservative scenario, with zero neutrino asymmetries, or as small as the baryon asymmetry, $\eta_B \sim 6 \cdot 10^{-10}$.
Finally, the last term in the righ-hand side of Eq.~(\ref{drhodt}) is the order $G_F^{2}$ collisional term.

The matter terms   can induce  Mikheyeev-Smirnov-Wolfenstein (MSW)-like resonances~\cite{Matt} 
when they become of the same order of the neutrino mass-squared splitting.
In the sterile sector resonances are associated with the three different
active-sterile mass splittings  $\Delta m^2_{4i}$ and with the different $\theta_{i4}$ mixing angles.
In particular,
the resonance condition can be satisfied (in both neutrino and antineutrino sectors) 
only for $\Delta m^2_{4i}<0$~\cite{Bell:1998ds,Dolgov:2001su,Mirizzi:2012we}. When more than one 
$\Delta m^2_{4i}$ is negative, multiple resonances can occur, affecting the sterile neutrino
production. Therefore, the resonance pattern is strongly dependent on the active and sterile
neutrino mass ordering  (see Fig.~\ref{fighier}).

\begin{figure}
\begin{center}
 \includegraphics[angle=0,width=0.5\textwidth]{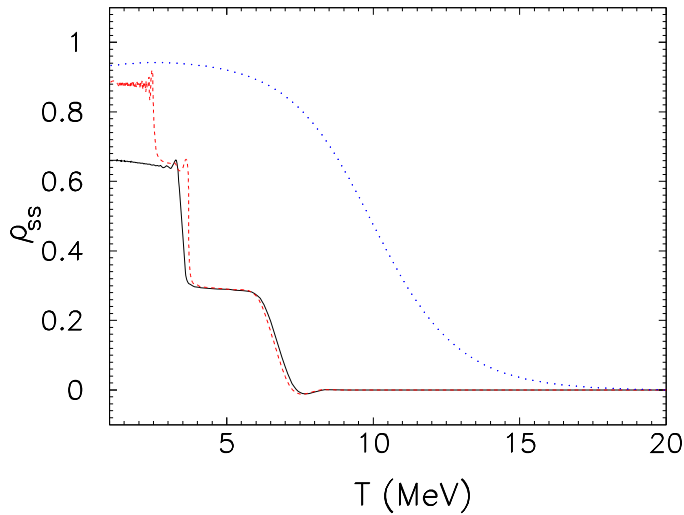} 
    \end{center}\label{fig1}
\caption{Evolution of the sterile neutrino density matrix $\rho_{ss}$ in the case
of $\sin^2\theta_{14}= 10^{-2}$, for $\Delta m^2_{41}= 10^{-5}$~eV$^{2}$ (continuous curve),  $\Delta m^2_{41}= -10^{-5}$~eV$^{2}$
(dashed curve) and $\Delta m^2_{41}=  5\times 10^{-2}$~eV$^{2}$ (dotted curve).
\label{resonance}} 
\end{figure}

An example of a possible resonance pattern is shown in Fig.~\ref{resonance}, where we represented the evolution as function
of the temperature $T$ of the sterile neutrino density matrix element $\rho_{ss}$ for $\sin^2\theta_{14}= 10^{-2}$
for different values of the  $\Delta m^2_{41}$ mass splitting.
In particular we take  three representative values, corresponding to different resonance schemes  of Fig.~1:  $\Delta m^2_{41}= 10^{-5}$~eV$^{2}$, corresponding to the NH-SNH scenario
with two resonances;  $\Delta m^2_{41}= -10^{-5}$~eV$^{2}$ corresponding to the NH-SIH scenario with three resonances;
$\Delta m^2_{41}= 5\times 10^{-2}$~eV$^{2}$ corresponding to the NH-SNH scenario with zero resonances.
 For
$\Delta m^2_{41}= 10^{-5}$~eV$^{2}$ (continuous curve) the two resonances are associated with $\Delta m^2_{43}$ and $\Delta m^2_{42}$, which  are negative. The first one is  at $T\simeq 6.5$~MeV between $\nu_4$ and $\nu_{2,3}$ associated with  $\Delta m^2_{43}$, and
the second one at  $T\simeq 3.5$~MeV between $\nu_4$ and $\nu_{1,2,3}$ associated with $\Delta m^2_{42}$.
For $\Delta m^2_{41}= -10^{-5}$~eV$^{2}$ (dashed curve) we also have a further resonance at $T\simeq 2.5$~MeV between
$\nu_4$ and $\nu_{1,2,3}$. Finally, for $\Delta m^2_{41}= 5\times 10^{-2}$~eV$^{2}$ since all the mass splitting between active and
sterile neutrinos are positive, there are no resonances.

 \section{Cosmological bounds on active-sterile $\nu$ mixing} 
 
Our cosmological bounds on sterile neutrinos are obtained exploiting the  Planck 95~\% C.L.  on $N_{\rm eff}$ and
 $\Omega_\nu h^2$ from  Eq.~(\ref{planckneff}). In terms of the neutrino density matrix these are given by
\begin{eqnarray}
N_{\rm eff} &=& \frac{1}{2} \textrm{Tr}[{\rho +{\bar\rho}}] \,\ , \nonumber \\
\Omega_\nu h^2 &=& \frac{1}{2} \frac{[\sqrt{\Delta m^2_{41}}({\rho_{ss} +{\bar\rho}_{ss}})]}{94.1 \,\  \textrm{eV}} \,\ .
\end{eqnarray}
The bound on $N_{\rm eff}$ extends also to small sterile neutrino masses, comparable to the ones given by the 
 two active mass splittings. 
However, since the   bounds on $N_{\rm eff}$ given by Planck are rather insensitive to  the active and sterile neutrino
 mass pattern~\cite{Planck2013}, we will refer to the value of  Eq.~(\ref{planckneff}) also in the cases of small sterile neutrino mass.
 The calculation of  $\Omega_\nu h^2 < 4.5 \times 10^{-3}$ assumes as dominant contribution the one given by
 the
only  sterile neutrino mass much larger than the active neutrino ones, 
 in agreement with the Planck analysis. Therefore, this bound is insensitive to the active
mass hierarchy.  
We  present our exclusion plots in the planes $(\Delta m^2_{41}, \sin^2 \theta_{i4})$.
Since the results as a function of $\sin^2 \theta_{34}$ and $\sin^2 \theta_{24}$ are very
similar, we omit to present the  $\sin^2 \theta_{34} \neq 0$  case. 
In Figure~\ref{NH} we consider NH, while in Figure~\ref{IH}
we refer to IH.
In each of these Figures,  the upper Panels a), b) refer to the the SNH case, while  the lower Panels 
c), d) are for
SIH case. Left panels refer to the exclusion plots in the plane $(\Delta m^2_{41}, \sin^2 \theta_{14})$ for 
different values of $ \sin^2 \theta_{24}$, while right panels refer to the plane 
 $(\Delta m^2_{41}, \sin^2 \theta_{24})$ for different values of $\sin^2 \theta_{14}$. 
In all  the cases,  $\sin^2 \theta_{34}$ is fixed to zero. The excluded regions from 
$N_{\rm eff}$   are those on the right or  
at the exterior of the  black contours, while the ones from $\Omega_\nu h^2$  are above the red contours.
We show the exclusion plots at  95 \% C.L.
   We now discuss the different panels of 
Figure~\ref{NH} and~\ref{IH} 
 in more details.


\subsection*{Active Normal hierarchy (Figure~\ref{NH})}
\subsubsection*{Sterile Normal hierarchy.}
 \emph{Panel a)}
We discuss at first the bound on $N_{\rm eff}$. 
 The most conservative limit corresponds to  $ \sin^2 \theta_{24}=0$, where for 
$\Delta m^2_{41}\gtrsim 10^{-2}$~eV$^{2}$ the exclusion contour is a straight line in this plane. 
We cut the bound at $\Delta m^2_{41} \sim 10^{-1}$~eV$^{2}$, 
since for higher masses 
sterile neutrinos would almost be non-relativistic at the CMB epoch, with a smaller contribution 
to the extra-radiation. 
However, as we will discuss later, this mass region can be strongly constrained using the bound coming from 
  $\Omega_\nu h^2$. As evident from Fig.~\ref{fighier},
lowering the value of $\Delta m^2_{41}$ one triggers  first a 
  $\nu_4-\nu_3$ resonance 
(when $\Delta m^2_{41} < \Delta m^2_{31}$) and then also a $\nu_4-\nu_2$ resonance 
(when $\Delta m^2_{41} < \Delta m^2_{21}$) which is the dominant one.
These produce the changes of the slope in the exclusion plot. 
Increasing the value of $\sin^2 \theta_{24}$, the constraint on the parameter space
becomes stronger. Large values of $ \sin^2 \theta_{24}$ would dominate the sterile neutrino
production and this excludes regions otherwise permitted if only $ \sin^2 \theta_{14}$ were non-zero. 
In particular, the only part that remains open is the transition region between the efficient non-resonant production range
at large $\Delta m^2_{41}$ and the one of resonant production at small $\Delta m^2_{41}$.
Finally, we represent  with a rectangle in the lower part of the plot, the region of parameters corresponding to a   light  sterile neutrino with $\Delta m^2_{41} \simeq 10^{-5}$~eV$^{2}$
and $\sin^2 \theta_{14} \sim 10^{-4}-10^{-3}$, suggested to solve the problem of the upturn of the solar neutrino spectrum~\cite{deHolanda:2010am}.
We realize that this region is excluded if  $\sin^2 \theta_{24} > 10^{-3}$. 

Passing now to the bounds from   $\Omega_\nu h^2$, again the most conservative limit is for
$ \sin^2 \theta_{24}=0$. In this case, values of $\Delta m^2_{41}\gtrsim 10^{-1}$~eV$^{2}$
are excluded for   $ \sin^2 \theta_{14} \gtrsim 10^{-2}$, since
the sterile neutrinos are fully thermalized. 
 For smaller values of 
 $\sin^2 \theta_{14} \gtrsim 10^{-2}$ the bound becomes less constraining, due to the incomplete
thermalization of the sterile species, allowing e.g.
  $\Delta m^2_{41}\lesssim 1$~eV$^{2}$ for $\sin^2 \theta_{14} \lesssim 10^{-4}$.
  The bound becomes more stringent increasing the value of $\sin^2 \theta_{24}$.
  In particular, for $\sin^2 \theta_{24} \gtrsim 10^{-2}$, $\Delta m^2_{41}\gtrsim 10^{-1}$~eV$^{2}$
  is excluded independently on the value of $\sin^2 \theta_{14}$, since sterile neutrinos would be
always produced with thermal abundances.

  For comparison, we also show the slice at
$\sin^2\theta_{24}=10^{-2}$
of the 95~\% C.L., for the allowed region
obtained from the global analysis
of short-baseline oscillation data
\cite{Giunti:2012tn,Archidiacono:2013xxa} (filled region in the up right part of the plot denoted by SBL). We observe
that it is  completely ruled out 
by the cosmological bound from $\Omega_\nu h^2$. We also plot the 90 \% C.L.  expected sensitivity of the KATRIN  experiment
(measuring the spectrum of electrons from tritium beta decay) after 3-years of data taking~\cite{Esmaili:2012vg}. 
Also this  region would be already excluded. 
 
  \emph{Panel b)} The description of the exclusion plot is analogous to the one of  Panel a), with the roles of $\theta_{14}$ and $\theta_{24}$ interchanged. In particular, the region of resonant
  sterile neutrino production is at $\Delta m^2_{41} \simeq 10^{-3}$~eV$^{2}$ when a   $\nu_4-\nu_3$ resonance
  is efficient. 
  Also the bound coming from  $\Omega_\nu h^2$ is comparable to the one shown in Panel a).

It is also shown 
 the slice at
$\sin^2\theta_{14}=10^{-1.5}$
of the 95~\% C.L. allowed region
obtained from the global analysis
of short-baseline oscillation data
\cite{Giunti:2012tn,Archidiacono:2013xxa},
which is another view of the SBL region shown in Panel a),
and the exclusion curve obtained from the combined analysis of
the data of $\nu_{\mu}$ and $\bar\nu_{\mu}$
disappearance experiments.
 One realizes that also in this case the region is 
excluded by the cosmological limit from  $\Omega_\nu h^2$.     
  
 \subsubsection*{Sterile Inverted hierarchy.}

 \emph{Panels c) and d)} We consider only 
 $|\Delta m^2_{41}| < 10^{-2}$~eV$^{2}$ due to the cosmological bound on active neutrino masses.
 Since  $\Delta m^2_{41} < 0$  an additional  $\nu_4$-$\nu_1$ resonance is present. This leads to an increase
in the production of sterile neutrinos, with respect to the SNH case. Therefore, the excluded regions in the parameter
space for the same values of the mixing angles are larger than the corresponding ones  in the upper panels.

\begin{figure*}
\begin{center}
 \includegraphics[width=0.4\textwidth]{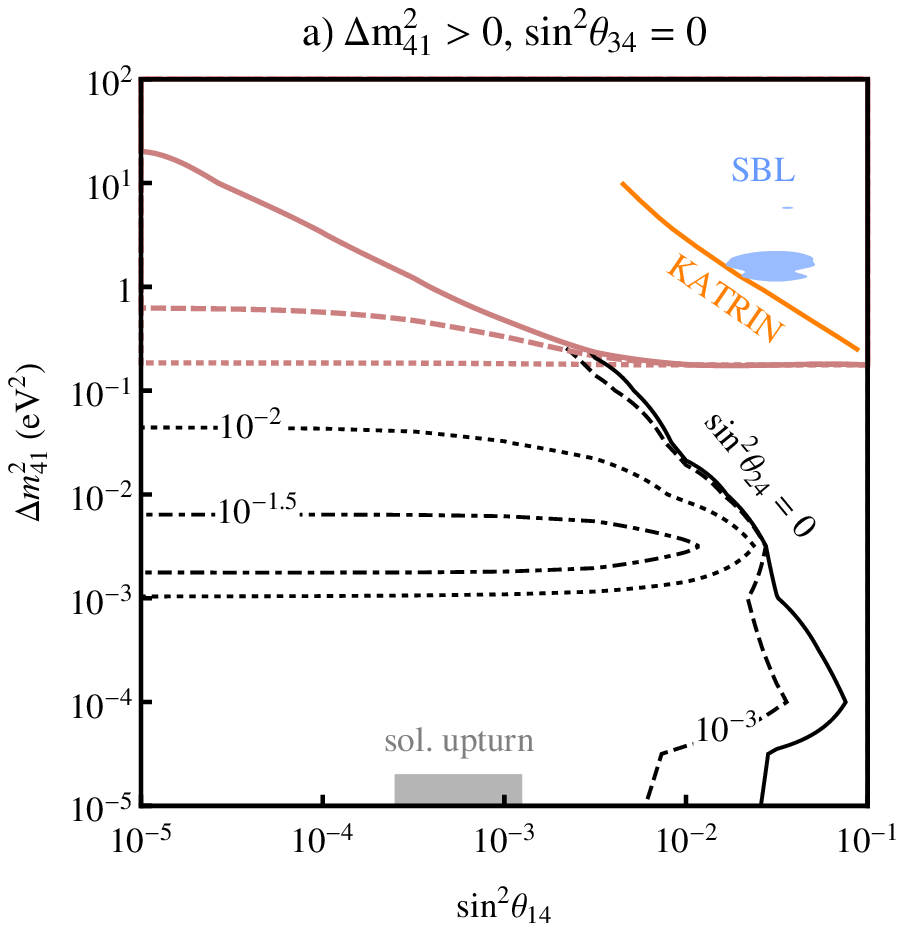} 
 \includegraphics[width=0.4\textwidth]{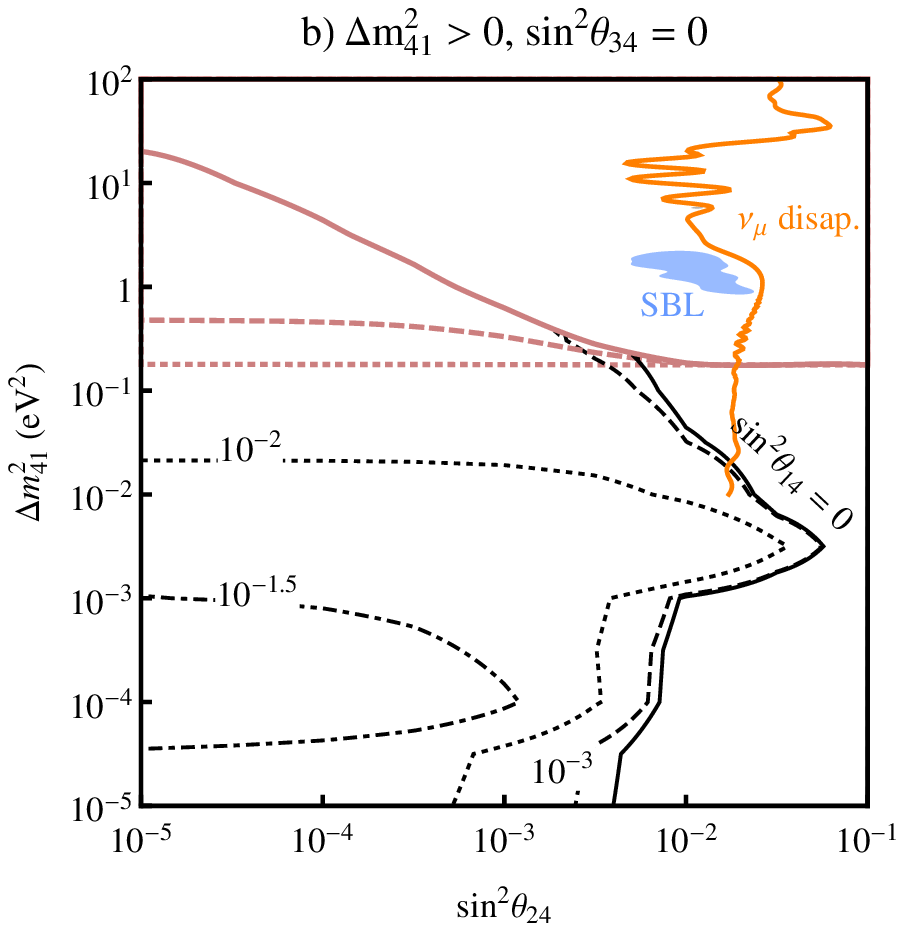} 
 \includegraphics[width=0.4\textwidth]{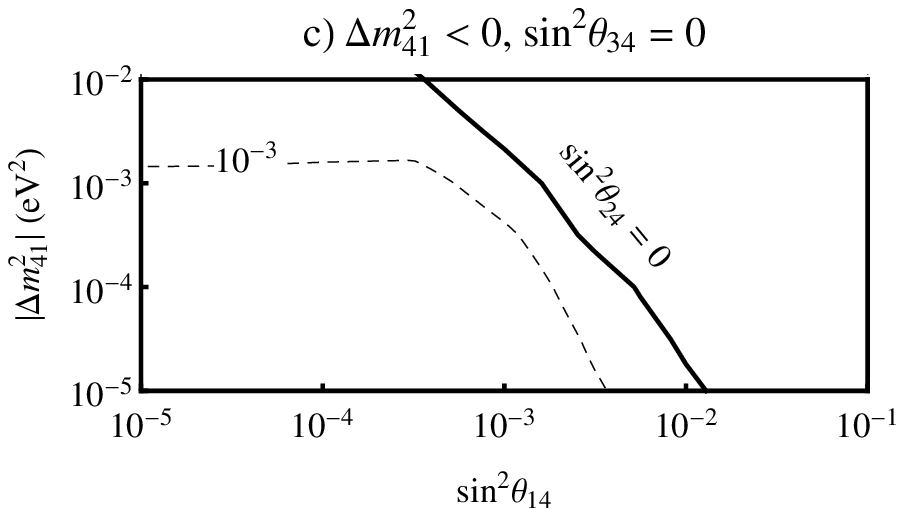} 
 \includegraphics[width=0.4\textwidth]{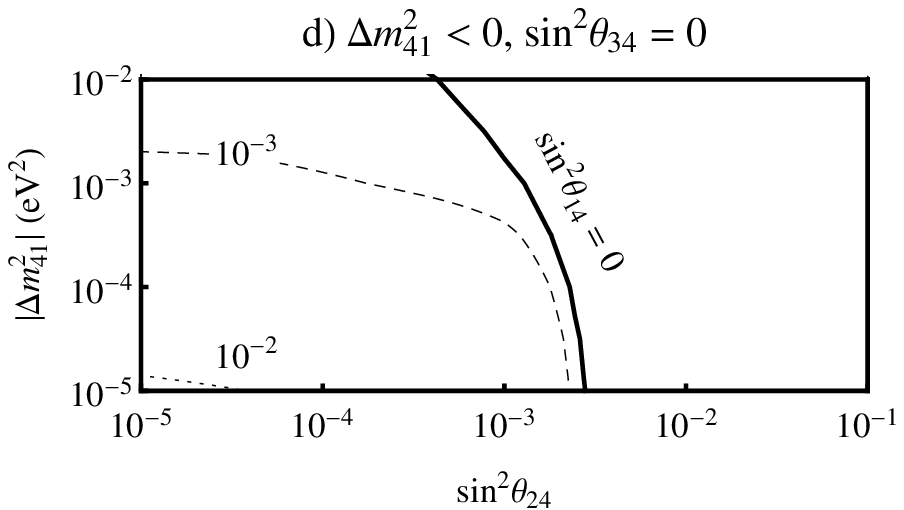} 
\end{center}
\caption{Active normal mass hierarchy NH. Exclusion plots for the active-sterile neutrino mixing parameter space for SNH
(upper panels) and SIH (lower panels) cases from $N_{\rm eff}$ (black curves) and  $\Omega_\nu h^2$ (red curves)   at  95 \% C.L. 
The contours refer to different values of $\sin^2 \theta_{i4}$: $\sin^2 \theta_{i4} =0$ (continuous curves),
$\sin^2 \theta_{i4} =10^{-3}$ (dashed curves), $\sin^2 \theta_{i4} =10^{-2}$ (dotted curves),
$\sin^2 \theta_{i4} =10^{-1.5}$ (dot-dashed curves). 
(see the text for details).} \label{NH}
\end{figure*}

\begin{figure*}
\begin{center}
 \includegraphics[width=0.4\textwidth]{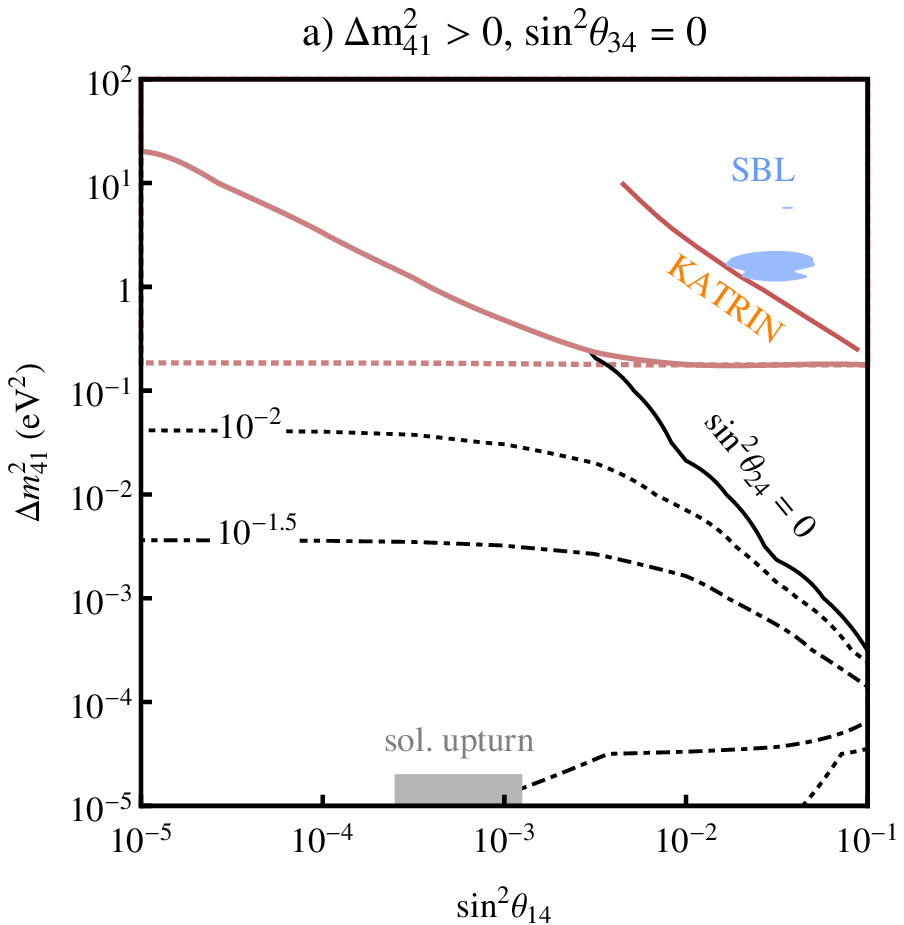} 
 \includegraphics[width=0.4\textwidth]{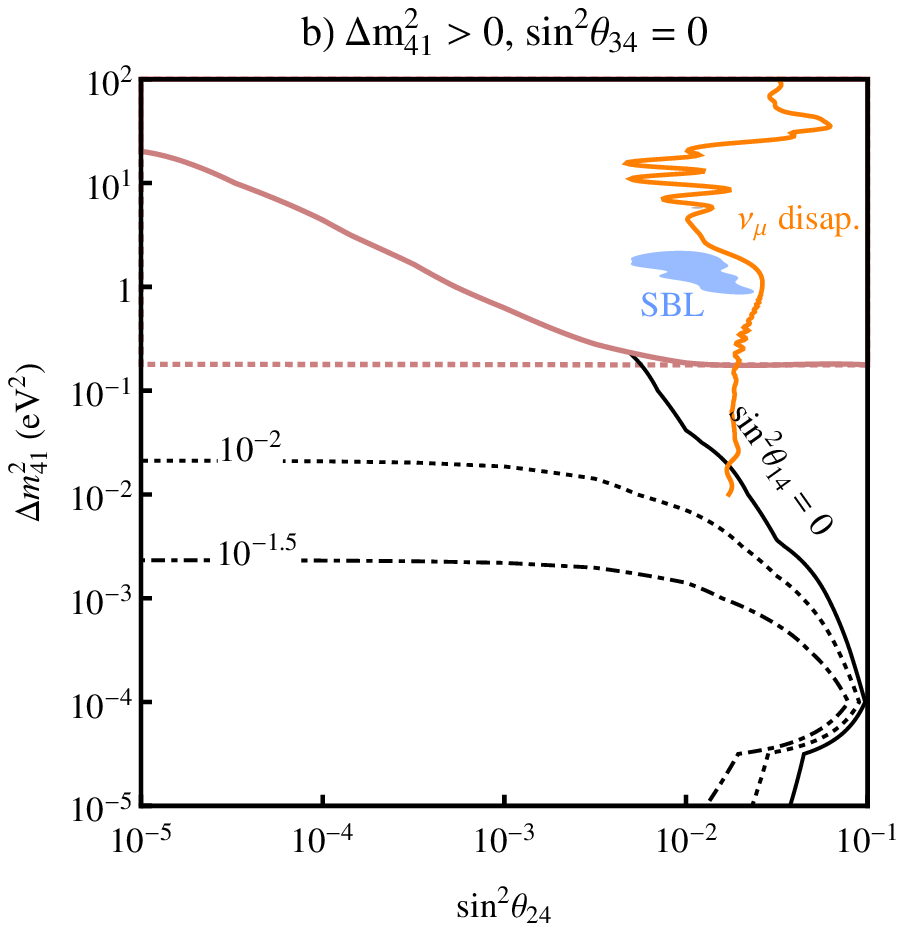} 
 \includegraphics[width=0.4\textwidth]{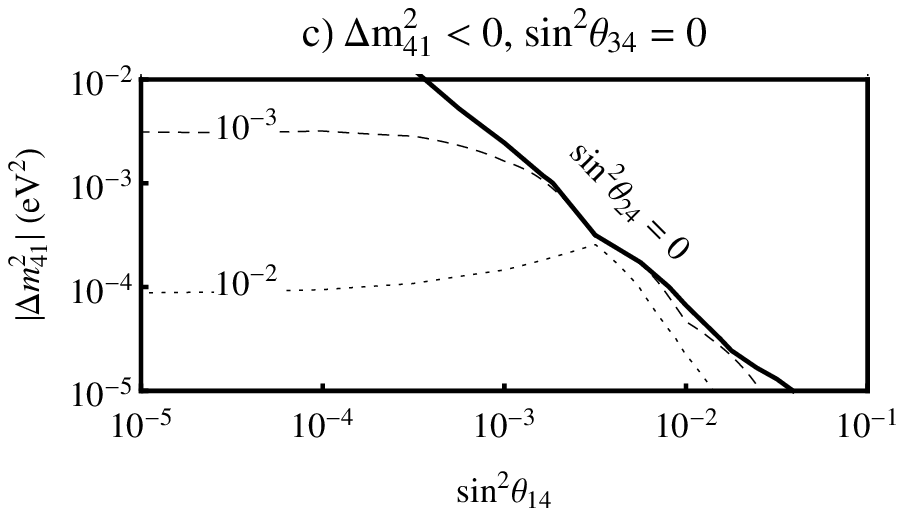} 
 \includegraphics[width=0.4\textwidth]{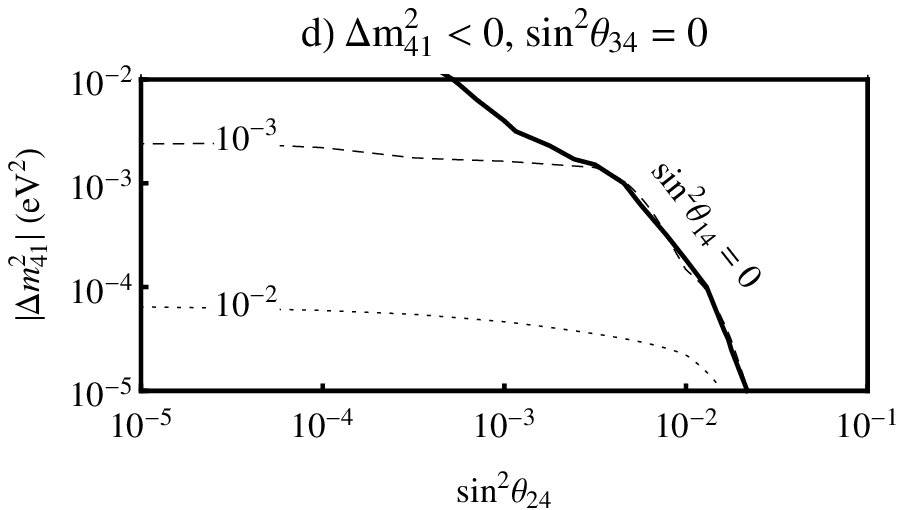} 
 \end{center}
\caption{Active inverted mass hierarchy IH. Exclusion plots for the active-sterile neutrino mixing parameter space for SNH
(upper panels) and SIH (lower panels) cases from $N_{\rm eff}$ (black curves)  and  $\Omega_\nu h^2$ (red curves)  at  95 \% C.L.
The contours refer to different values of $\sin^2 \theta_{i4}$: $\sin^2 \theta_{i4} =0$ (continuous curves),
$\sin^2 \theta_{i4} =10^{-2}$ (dotted curves),
$\sin^2 \theta_{i4} =10^{-1.5}$ (dot-dashed curves). 
 (see the text for details).} \label{IH}
\end{figure*}
 
\subsection*{Active Inverted hierarchy (Figure~\ref{IH})} 
 
 \subsubsection*{Sterile Normal hierarchy.}

 \emph{Panels a) and b)} From Fig.~\ref{fighier} it results that there can be only 
a single resonance for
 $\Delta m^2_{41} < \Delta m^2_{21}$. Therefore, comparing the exclusion plots from 
$N_{\rm eff}$
with the corresponding ones in
 Fig.~\ref{NH} one realizes that the constraint is less stringent. In particular, in Panel b)  the change
 in the slope in the exclusion plot is at $\Delta m^2_{41} \sim \Delta m^2_{21} \sim 10^{-4}$~eV$^{2}$, i.e.
at a smaller value with respect to   Fig.~\ref{NH}.
Concerning the bound from $\Omega_\nu h^2$, since it occurs in a region where $\Delta m^2_{41}$ is much larger than
the active mass splittings, it is independent on the mass hierarchy and so it is the same as in Fig.~\ref{NH}.

 \subsubsection*{Sterile Inverted hierarchy.}

\emph{Panels c) and d)} In this case, looking at Fig.~\ref{fighier} we realize that 
for $|\Delta m^2_{41}| > |\Delta m^2_{31}|$  three resonances are possible as in the NH case shown in the Fig.~\ref{NH},
while for  $|\Delta m^2_{41}| < |\Delta m^2_{31}|$ only two resonances  occur. Therefore, for 
 $|\Delta m^2_{41}| \lesssim 10^{-4}$~eV$^{2}$ the constraint from $N_{\rm eff}$ becomes less stringent than
 in the corresponding case in the NH scenario, while it is  comparable for larger mass splittings.

\section{Conclusions}    
In this paper we have exploited the very recent measurement of $N_{\rm eff}$ and
 $\Omega_\nu h^2$ provided by the Planck experiment  to update 
 the cosmological bounds on (3+1) sterile neutrino scenarios under the assumption of vanishing or very small neutrino asymmetries, of the order of the baryonic one. At this regard, for the first time it is shown how the constraints change if two  active-sterile mixing angles are considered.
 
We find that the sterile neutrino parameter space is severely constrained, and the excluded area
from the bound on $\Omega_\nu h^2$  covers the region accessible by current and future laboratory experiments.   
Moreover, from the results of our analysis we conclude that there is a  tension with the sterile neutrino hints from short-baseline experiments.
In particular, in the scenario we considered sterile neutrinos with $m\sim {\mathcal O}(1)$~eV would be excluded 
at more than $4$-$\sigma$. 
 Notice that combining Planck findings with other data might further strengthen the bounds on $N_{\rm eff}$. For example, adding to the analysis the primordial deuterium determination of Ref.~\cite{Pettini:2012ph}, compared with the BBN theoretical expectation as function of baryon density and $N_{\rm eff}$, leads to $N_{\rm eff} \leq 3.56$ at 95~\% C.L. \cite{Planck2013}. This means that future $^2$H measurements reducing the present spread of different Quasar Absorption System results  would lead to stronger bounds on sterile neutrino mixing parameters.

In order to reconcile the laboratory signals in favor of extra sterile neutrino degrees of freedom with the cosmological bounds one should introduce some extra parameters in the so far extremely succesful standard cosmological model, as for example, large neutrino--antineutrino asymmetries,
 $L_{\nu} = (n_\nu -n_{\bar \nu})/n_\gamma \gtrsim 10^{-2}$~\cite{Hannestad:2012ky, Mirizzi:2012we, Saviano:2013ktj}, which might inhibit the sterile neutrino production in the early universe.
After all, the fact that a completely satisfactory model of everything might not yet achieved is welcome, as it would continue to trigger the curiosity of physicists to look for what is, hopefully, beyond the corner.
 
\vspace{-0.5cm} 
 
\section*{Acknowledgements} 
We warmly thank Pasquale Dario Serpico for stimulating this project with useful discussions.
A.M. and N.S. thank Jan Hamann for useful discussions on Planck data.
The work of E. B., A.M. and N.S.  was supported by the German Science Foundation (DFG)
within the Collaborative Research Center 676 ``Particles, Strings and the
Early Universe''. C.G., G. M., G. M., and O.P.  acknowledge support by
the {\it Istituto Nazionale di Fisica Nucleare} I.S. FA51. 


\end{document}